\documentclass[pre,twocolumn,showpacs,amsmath,amssymb]{revtex4}
\usepackage[dvipdfmx]{graphicx}
\usepackage{epstopdf}
\usepackage{dcolumn}
\usepackage{bm}
\usepackage{booktabs}

\begin{document}

\title{Nonequilibrium behaviors of 3D Heisenberg model 
in the Swendsen-Wang algorithm}

\author{Yoshihiko Nonomura}
\email{nonomura.yoshihiko@nims.go.jp}
\affiliation{Computational Materials Science Unit, 
National Institute for Materials Science, Tsukuba, Ibaraki 305-0044, Japan} 

\author{Yusuke Tomita}
\email{ytomita@shibaura-it.ac.jp}
\affiliation{College of Engineering, Shibaura Institute of Technology, 
Saitama 337-8570, Japan} 

\begin{abstract}
Recently Y.~N.\ showed that the nonequilibrium critical relaxation of the 
2D Ising model from the perfectly-ordered state in the Wolff algorithm is 
described by the stretched-exponential decay, and found a universal 
scaling scheme to connect nonequilibrium and equilibrium behaviors. 
In the present study we extend these findings to vector spin models, 
and the 3D Heisenberg model could be a typical example. 
In order to evaluate the critical temperature and critical exponents 
precisely with the above scaling scheme, we calculate the 
nonequilibrium ordering from the perfectly-disordered state in the 
Swendsen-Wamg algorithm, and find that the critical ordering 
process is described by the stretched-exponential growth with 
the comparable exponent to that of the 3D XY model. 
The critical exponents evaluated in the present study 
are consistent with those in previous studies.
\end{abstract}


\maketitle

\section{Introduction}
The cluster algorithm was first proposed~\cite{SW} 
in the Potts model~\cite{PottsRev}, which is a generalized 
version of the Ising model and characterized by short-range 
interactions between discrete local variables. The partition 
function of this model can be represented with respect to 
percolation clusters~\cite{FK}, and the cluster algorithm 
is formulated on the basis of this representation. 
In this algorithm temperature is included in the probability 
for connecting local variables to construct the clusters, and 
the flipping rate of clusters does not depend on temperature. 
Then, this algorithm has been expected to reduce the critical 
slowing down in the vicinity of the critical temperature $T_{\rm c}$, 
which is characterized by the dynamical critical exponent $z$ 
defined by divergence of the relaxation time of autocorrelation 
functions as $\tau(T) \sim (T-T_{\rm c})^{-z}$. 

In finite systems the relaxation time is finite even at 
$T_{\rm c}$, and the exponent $z$ is estimated from 
its size dependence as $\tau(T_{\rm c}, L) \sim L^{z}$. 
Large reduction of $z$ from that in the local-update 
algorithms was exhibited even in the pioneering 
articles of the cluster algorithms~\cite{SW,Wolff}. 
In some numerical calculations it was pointed out that 
the weak power-law and logarithmic size dependences of 
$\tau(L)$ were difficult to distinguish~\cite{Heermann,Baillie}.  
Such logarithmic size dependence stands for $z=0$, namely 
the power-law dynamical critical behaviors may be questionable 
in the cluster algorithms. Then, much larger-scale calculations 
were performed by G\"und\"u\c{c} {\it et al.}~\cite{Gunduc} and 
Du {\it et al.}~\cite{Du} with the nonequilibrium relaxation (NER), 
and they found that $z=0$ widely holds in the cluster algorithms. 

Recently one of the present authors (Y.~N.) 
found~\cite{Nonomura14} that the explicit form of 
NER of the magnetization in the 2D Ising model with the 
Wolff algorithm from the perfectly-ordered state is described 
by the stretched-exponential time dependence, and proposed 
a universal scaling scheme in which the whole relaxation process 
from early-time nonequilibrium behaviors to equilibrium ones 
for various system sizes is located on a single curve.
In the present article we generalize these findings to 
vector spin models, and choose the 3D Heisenberg model as a 
typical example, since it shows the second-order phase transition 
and the critical behaviors have already been studied intensively, 
similarly to the Ising models. 

Outline of the present article is as follows: In Section II, formulation 
based on the ``embedded-Ising-spin" algorithm is briefly reviewed, 
and the reason why our calculations is based on the ordering from 
the perfectly-disordered state in the Swendsen-Wang (SW) algorithm 
is explained. In Section III, explicit scaling forms of various physical 
quantities with size-dependent factors are exhibited. The critical 
temperature $T_{\rm c}$ is evaluated from the ordering process 
of magnetization together with the critical exponent $\beta/\nu$, and 
other exponents $\gamma/\nu$ and $\alpha/\nu$ are estimated from 
the scaling behaviors of the magnetic susceptibility and specific heat. 
Then, the exponent $\nu$ can be obtained from the scaling relation. 
However, evaluation of $\alpha/\nu$ from the specific heat is difficult 
in 3D vector spin models, and $\nu$ is also estimated from the 
scaling behavior of the temperature derivative of magnetization. 
In section IV, numerical results of the present articles are discussed. 
Estimates of the critical exponents $\beta$, $\gamma$ and $\nu$ 
are compared with other numerical studies, and promising 
future tasks along the present framework are mentioned. 
In section V, the above descriptions are summarized. 
Evaluation of $T_{\rm c}$ from single system and 
nonequilibrium relaxation from the perfectly-ordered state 
in the SW and Wolff algorithms are treated in Appendices. 

\section{Formulation}
In the present article, Monte Carlo simulations of the 
ferromagnetic 3D Heisenberg model on a cubic lattice,
\begin{equation}
\label{Hei-Ham}
{\cal H}=-J\sum_{\langle ij \rangle \in {\rm n.n.}}\vec{S}_{i} \cdot \vec{S}_{j}
\end{equation}
with summation over all the nearest-neighbor bonds, 
are performed with the cluster algorithm. Wolff~\cite{Wolff} 
showed that the cluster update of vector spins such as 
Heisenberg one is possible by constructing spin clusters with 
respect to the Ising element projected onto a randomly-chosen 
direction $\vec{r}$, $|\vec{r}|=1$ in each Monte Carlo step. 
That is, two nearest-neighbor spins $\vec{S}_{i}$ and $\vec{S}_{j}$ 
are connected with probability 
$p=1-\exp [ -2 \beta  (\vec{r} \cdot \vec{S}_{i}) (\vec{r} \cdot \vec{S}_{j}) ]$ 
with $\beta \equiv J/T$, 
when the projected Ising elements along $\vec{r}$, namely 
$(\vec{r} \cdot \vec{S}_{i})\vec{r}$ and  $(\vec{r} \cdot \vec{S}_{j})\vec{r}$, 
are aligned along the same direction. 

Wolff proposed~\cite{Wolff} the so-called Wolff algorithm 
in the same article, in which a single spin cluster 
generated from a randomly-chosen spin is always flipped. 
On the other hand, even in Wollff's  ``embedded-Ising-spin" 
scheme, the SW update~\cite{SW} is also possible, 
in which all spins are swept in every step and clusters with 
the flipping rate $50\%$ are generated in the whole system. 
Although both approaches result in the same equilibrium, 
dynamical process of them might be different. 
Actually, even in the Ising model, dynamics of these two 
algorithms are different in the ordering process from 
the perfectly-disordered state~\cite{Du}. 
While physical quantities show stretched-exponential 
behaviors in the SW algorithm, they show power-law 
behaviors in the Wolff algorithm. 

While the cluster to be grown to system-size one is 
swept in each step in the SW algorithm by definition, 
such an event is quite rare in the Wolff algorithm. 
Since calculations from the perfectly-disordered state 
is required for the NER analysis of physical quantities 
with respect to fluctuations~\cite{Ito00,NERrev}, 
we utilize the SW algorithm in the present article. 
Furthermore, dynamical process in these two cases are 
different even in calculations from the perfectly-ordered 
state in the 3D Heisenberg model, which will be explained 
in Appendix B~\cite{app-order}.

\section{Numerical results}
\subsection{Overview of calculations}
Recently one of the present authors (Y.~N.) found~\cite{Nonomura14} 
that nonequilibrium critical relaxation of the absolute value of 
magnetization in the 2D Ising model is described by the 
stretched-exponential decay, 
\begin{equation}
\label{se-ord}
\langle |m(t)| \rangle \sim \exp \left[ - ( t / \tau'_{m} )^{\sigma} \right],\ 
0<\sigma <1,
\end{equation}
in the Wolff algorithm from the perfectly-ordered state. 
This scaling form also holds in the SW algorithm. 
When similar calculations are started from the 
perfectly-disordered state, the power-law ordering 
is observed in the Wolff algorithm~\cite{Du}. 
On the other hand, such ordering is described by 
the stretched-exponential one, 
\begin{equation}
\label{se-dis}
\langle |m(t,L)| \rangle \sim L^{-d/2} \exp \left[ + ( t / \tau_{m} )^{\sigma} \right],\ 
0<\sigma <1,
\end{equation}
in the SW algorithm~\cite{Nono-full}. Here early-time size 
dependence of the magnetization is also taken into account, 
and this form is derived from normalized random-walk growth 
of clusters. In the present article we also analyze other 
physical quantities on the basis of similar scaling forms. 
In order to investigate such diverging quantities at $T_{\rm c}$, 
calculations from the perfectly-disordered state are 
required in the NER method. Since the power-law ordering 
in the Wolff algorithm has no essential difference from 
that in the local-update algorithm, here we concentrate 
on calculations based on the SW algorithm.

The purpose to treat some physical quantities are to evaluate critical 
exponents. Although they do not appear explicitly in Eq.\ (\ref{se-dis}), 
the magnetization at the critical temperature $T_{\rm c}$ converges to 
the critical one scaled as $m_{\rm c}(L) \sim L^{-\beta/\nu}$ in equilibrium. 
Then, multiplying $L^{\beta/\nu}$ for both sides of Eq.\ (\ref{se-dis}) and 
taking logarithm, we have 
\begin{equation}
\label{eq-scale}
\log (\langle |m(t,L)| \rangle L^{\beta/\nu})
\sim 
(t / \tau_{m})^{\sigma} -\log L^{d/2-\beta/\nu}.
\end{equation}
When lhs of Eq.\ (\ref{eq-scale}) is plotted versus rhs of it 
for several system sizes, short-time (in the region Eq.\ (\ref{se-dis}) holds) 
and long-time (in the vicinity of equilibrium) data are scaled by definition. 
Actually, even the data between these two regions are also scaled on a 
single curve~\cite{Nonomura14,Nono-full}, and the exponent $\beta/\nu$ 
can be evaluated from this ``nonequilibrium-to-equilibrium" scaling plot. 
Although this plot seems to have 3 parameters $\sigma$, $\beta/\nu$ 
and $\tau_{m}$, fitting procedure is not so difficult. 
The data close to equilibrium almost only depend on $\beta/\nu$, and 
there exists a constraint that the tangent of the short-time data is unity. 
Other exponents $\gamma/\nu$ and $\alpha/\nu$ can be estimated from 
similar scaling plots of the magnetic susceptibility and specific heat. 
Then, the exponent $\nu$ can be obtained from 
the hyperscaling relation, $2-\alpha = d \nu$.

However, direct evaluation of $\alpha/\nu$ from size dependence of 
the specific heat is not trivial in the 3D vector spin models, because 
the critical exponent $\alpha$ is known to be close to zero or even 
negative in these models. In such a case, the temperature derivative of 
magnetization~\cite{Schulke96,Ito00,NERrev} might be a good quantity. 
Since the spontaneous magnetization behaves as 
$m_{\rm s}\sim (T_{\rm c}-T)^{\beta}$ in the vicinity of $T_{\rm c}$, 
its temperature derivative behaves as 
${\rm d}m_{\rm s}/{\rm d}T \sim -(T_{\rm c}-T)^{\beta-1}$, and 
the size dependence of this quantity at $T_{\rm c}$ is given by 
${\rm d}m_{\rm s}(L; T_{\rm c})/{\rm d}T \sim - L^{(1-\beta)/\nu}$. 

In order to evaluate the critical exponents precisely enough, the critical 
temperature $T_{\rm c}$ should also be evaluated precisely beforehand. 
In the traditional NER scheme based on the local-update algorithm, 
$T_{\rm c}$ is evaluated from linearity of the log-log plot of relaxation 
data for a single system size in the power-law region. This region 
becomes fairly wide when rather large systems are considered. 
Usually, saturation to equilibrium behaviors is never observed 
in the standard NER analysis. 
On the other hand, the stretched-exponential region is rather 
narrow in the NER analysis based on the cluster algorithms. 
Especially in three or more dimensions, precise evaluation 
of $T_{\rm c}$ from a single system size is challenged by 
saturation to equilibrium behaviors. Such attempts will be 
summarized in Appendix A~\cite{app-order}, and the final-stage 
estimation of $T_{\rm c}$ coupled with determination of 
critical exponents is explained in the present section.

\subsection{Evaluation of $T_{\rm c}$ and $\beta/\nu$ from magnetization}
\begin{figure}
\includegraphics[width=88mm]{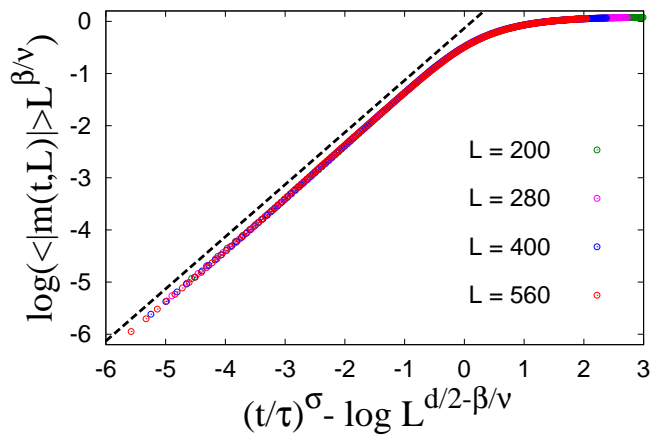}
\caption{(Color online) Nonequilibrium-to-equilibrium scaling 
plot of magnetization at $T_{\rm c}=1.442987 J/k_{\rm B}$ for 
$L=200$, $280$, $400$ and $560$ with $\beta/\nu=0.516$, 
$\sigma=0.47$ and $\tau_{m}=2.51$. 
All the data seem to be located on a single curve within this scale. 
The dashed line stands for the tangent $1$ as guides for eyes.
}
\label{fig-mag-sc}
\end{figure}
\begin{figure}
\includegraphics[width=88mm]{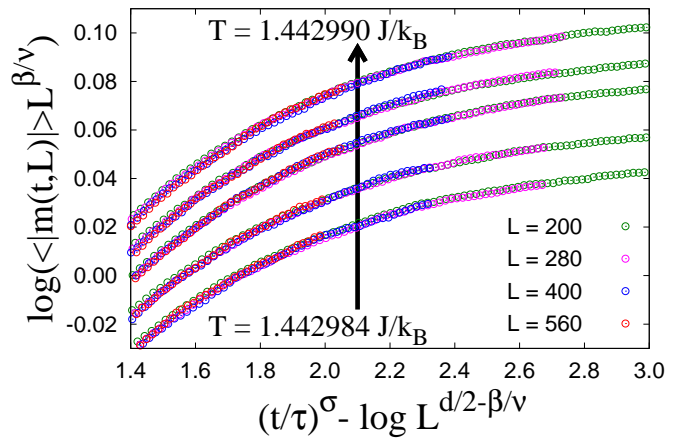}
\caption{(Color online) Enlarged plot in the vicinity 
of equilibrium corresponding to Fig.\ \ref{fig-mag-sc}  at 
$T=1.442984 J/k_{\rm B}$, $1.442985 J/k_{\rm B}$, 
$1.442987 J/k_{\rm B}$, $1.442989 J/k_{\rm B}$ and 
$1.442990 J/k_{\rm B}$ with $\beta/\nu=0.511$, $0.512$, 
$0.516$, $0.518$ and $0.519$, respectively. 
Error bars are smaller than symbols.}
\label{fig-mag-eq}
\end{figure}

First, the scaling plot of magnetization (\ref{eq-scale}) at the 
estimated critical temperature $T_{\rm c}=1.442987 J/k_{\rm B}$ 
(to be explained later) with $\beta/\nu=0.516$, $\sigma=0.47$ 
and $\tau_{m}=2.51$ is displayed in Fig.\ \ref{fig-mag-sc}, where 
all the data points for $L=200$ (averaged with $320,000$ 
random number sequences (RNS)), $280$ ($160,000$ RNS), 
$400$ ($80,000$ RNS) and $560$ ($80,000$ RNS) 
seems to be located on a single curve, as expected. 
Evaluation process of $T_{\rm c}$ and $\beta/\nu$ 
is shown in Fig.\ \ref{fig-mag-eq}, where the data for 
$T=1.442984 J/k_{\rm B}$, $1.442985 J/k_{\rm B}$, 
$1.442987 J/k_{\rm B}$, $1.442989 J/k_{\rm B}$ and 
$1.442990 J/k_{\rm B}$ (from bottom to top) in the saturation 
region are plotted with Eq.\ (\ref{eq-scale}), and the data are 
almost on a single curve after the tuning of parameters. 
The finest scaling is observed at $T=1.442987 J/k_{\rm B}$ 
with $\beta/\nu=0.516$, while deviations of the data at 
$T=1.442984 J/k_{\rm B}$ with $\beta/\nu=0.511$ and 
at $T=1.442990 J/k_{\rm B}$ with $\beta/\nu=0.519$ 
are not negligible. Since the direction of deviations 
is not systematic as varying the system size, 
these deviations cannot be reduced anymore. 
We take similar scaling plots for 
$1.442985 J/k_{\rm B} \leq T \leq 1.442989 J/k_{\rm B}$, 
and observe comparable scaling behaviors with 
gradually-varying exponent $0.512 \leq \beta/\nu \leq 0.518$. 
The fine scaling behaviors for 
$T=1.442985 J/k_{\rm B} \leq T \leq 1.442989 J/k_{\rm B}$ 
becomes as bad as that at $T=1.442984 J/k_{\rm B}$ by 
changing $\beta/\nu$ with $\pm 0.002$. 
Combining these results, we can safely conclude 
\begin{equation}
\label{Tc-est}
T_{\rm c}=1.442987(2) J/k_{\rm B}, \ \beta/\nu=0.515 \pm 0.005.
\end{equation}

When we fit the data for $L=560$ at $T_{\rm c}$ directly with 
Eq.\ (\ref{se-dis}), we have $\sigma \approx 0.49$, which is 
consistent with that of the 3D XY model~\cite{Nono-letter} 
and the exponent $\sigma \approx 1/2$ might be common 
in all the vector spin models characterized by the 
second-order phase transition. 
In this fitting the initial $\sim 50$ MCS data are used 
by minimizing the residue per data~\cite{Nonomura14}. 
With this exponent the data between the stretched-exponential 
and equilibrium regions are not scaled so well as those plotted 
in Fig.\ \ref{fig-mag-sc} due to higher-order correction of 
simulation time, and the deviation from $\sigma=1/2$ may 
compensate such higher-order correction numerically.

\begin{figure}
\includegraphics[width=88mm]{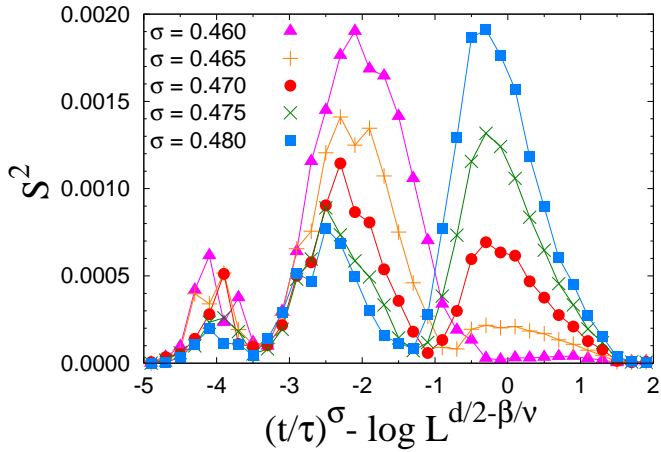}
\caption{(Color online) Average mutual variance of the scaling 
plot in Fig.\ \ref{fig-mag-sc} for $\sigma=0.460$ (triangles), 
$0.465$ (crosses), $0.470$ (circles), $0.475$ (x-marks) 
and $0.480$ (squares). Overall variance seems to be 
the smallest for $\sigma \approx 0.47$. 
}
\label{fig-mag-sig}
\end{figure}
Evaluation process of $\sigma$ is given in Fig.\ \ref{fig-mag-sig}, 
which exhibits the average mutual variance $S^{2}$ in the scaling 
plot like Fig.\ \ref{fig-mag-sc} at $T_{\rm c}=1.442987 J/k_{\rm B}$ 
for $\sigma=0.460$ (triangles), $0.465$ (crosses), 
$0.470$ (circles), $0.475$ (x-marks) and $0.480$ (squares). 
After the optimization of parameters $\beta/\nu$ and $\tau_{m}$ with 
given $\sigma$ for $L=200$, $280$, $400$ and $560$, variance 
of the data between different $L$ at the same scaled simulation 
time $t_{\ast} \equiv (t/\tau_{m})^{\sigma}- \log L^{d/2-\beta/\nu}$ 
(between the law data for one size and the linearly-interpolated 
ones for other sizes) is averaged in the width $\Delta t_{\ast}=0.2$ 
(for example, from $t_{\ast}=-5.0$ to $-4.8$). 

Apparently, the variance is very small at the onset of ordering 
($t_{\ast} \approx -5$) and in the vicinity of equilibrium 
($t_{\ast} \approx 2$), and has two main peaks on the edge 
of the stretched-exponential ordering ($-3 < t_{\ast} < -1$; 
region A) and between the stretched-exponential and 
equilibrium regions ($-1 < t_{\ast} < 1$; region B). 
For $\sigma=0.48$, the variance is the smallest in the 
region A while the largest in the region B, and {\it vice versa} 
for $\sigma=0.46$. For $\sigma=0.47$ the variance takes 
medium values in the both regions and these two values 
are almost the same, and therefore the most suitable 
within these three values. Behaviors for $\sigma=0.475$ 
and $0.465$ just locate in between. Optimized parameters 
are $\beta/\nu=0.515$ and $\tau_{m}=2.31$ for $\sigma=0.46$, 
and $\beta/\nu=0.517$ and $\tau_{m}=2.74$ for $\sigma=0.48$. 
That is, the variance of $\beta/\nu$ is much smaller than that 
of $\sigma$. 

Calculations for the variance of the next order of $\sigma$ 
are affected by the precision of optimization of parameters 
or the method of interpolation (linear or up to higher orders). 
Such efforts may not be productive, because error bars in 
$\beta/\nu$ are smaller than those from the scaling analysis 
in Fig.\ \ref{fig-mag-eq} ($\pm 0.002$) for each temperature. 
Then, here we evaluate this exponent as $\sigma=0.47(1)$ 
and use this estimate for analyses of other physical quantities.

\subsection{Evaluation of $\gamma/\nu$ from magnetic susceptibility}
\begin{figure}
\includegraphics[width=88mm]{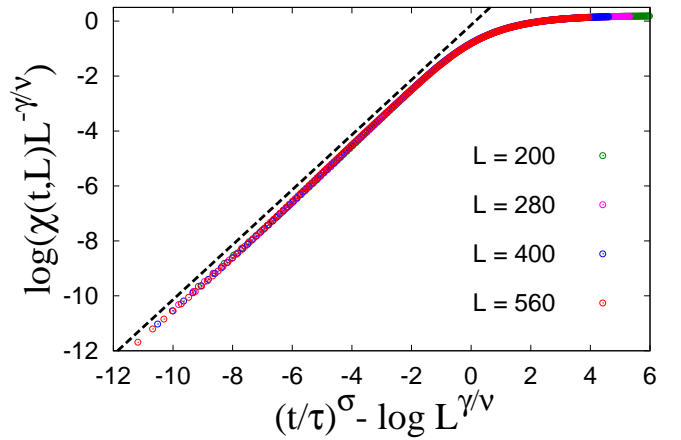}
\caption{(Color online) Nonequilibrium-to-equilibrium scaling 
plot of magnetic susceptibility at $T_{\rm c}=1.442987 J/k_{\rm B}$ 
for $L=200$, $280$, $400$ and $560$ with $\gamma/\nu=1.970$, 
$\sigma=0.47$ and $\tau_{\chi}=0.584$. 
All the data seem to be located on a single curve, and the 
dashed line stands for the tangent $1$ as guides for eyes.
}
\label{fig-chi-sc}
\end{figure}
\begin{figure}
\includegraphics[width=88mm]{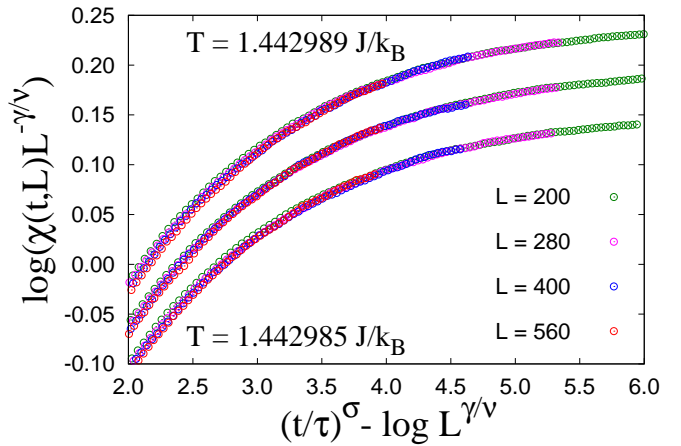}
\caption{(Color online) Enlarged plot in the vicinity 
of equilibrium corresponding to Fig.\ \ref{fig-chi-sc}  at 
$T=1.442985 J/k_{\rm B}$, $1.442987 J/k_{\rm B}$ 
and $1.442989 J/k_{\rm B}$ with $\gamma/\nu=1.977$, 
$1.970$ and $1.967$, respectively. 
Error bars are smaller than symbols.
}
\label{fig-chi-eq}
\end{figure}

Although evaluation of $T_{\rm c}$ based on the magnetic susceptibility 
is also possible, here we use $T_{\rm c}$ based on magnetization 
and estimate $\gamma/\nu$ similarly to the scheme used in 
Fig.\ \ref{fig-mag-eq}. First, the magnetic susceptibility at $T_{\rm c}$ 
in the SW algorithm shows the stretched-exponential ordering, 
\begin{equation}
\chi(t,L)
\equiv \frac{1}{L^{d}} \sum_{i,j}
           \langle \vec{S}_{i} \cdot \vec{S}_{j} \rangle
\sim \exp \left[ + ( t / \tau_{\chi} )^{\sigma} \right]. 
\end{equation}
This physical quantity has no size-dependent prefactor because 
duplicated random-walk growth of clusters ($\sim L^{-(d/2)\times 2}$) 
is cancelled with bulk response ($\sim L^{d}$). Since equilibrium 
value of this quantity is scaled as $\chi_{\rm c}\sim L^{\gamma/\nu}$, 
we have the following scaling form similar to Eq.\ (\ref{eq-scale}), 
\begin{equation}
\log ( \chi(t,L) L^{-\gamma/\nu})
\sim 
(t / \tau_{\chi})^{\sigma} -\log L^{\gamma/\nu}.
\end{equation}
The scaling plot at $T_{\rm c}=1.442987 J/k_{\rm B}$ with 
$\gamma/\nu=1.970$, $\sigma=0.47$ and $\tau_{\chi}=0.584$ 
is displayed in Fig.\ \ref{fig-chi-sc}, where all the data points for 
$L=200$, $280$, $400$ and $560$ seems to be located 
on a single curve again. The data in the vicinity of equilibrium 
is shown in Fig.\ \ref{fig-chi-eq}, together with the data for 
$T=1.442985 J/k_{\rm B}$ (with $\gamma/\nu=1.977$) 
and $1.442989 J/k_{\rm B}$ (with $\gamma/\nu=1.967$). 
Discrepancy from such scaling behaviors becomes 
clear by changing $\gamma/\nu$ with $\pm 0.002$. 
Combining these results, we conclude 
\begin{equation}
\label{gamma-est}
\gamma/\nu=1.972 \pm 0.007.
\end{equation}

\subsection{Evaluation of $\nu$ from temperature derivative 
of magnetization and $\alpha$ from scaling relation}
\begin{figure}
\includegraphics[width=88mm]{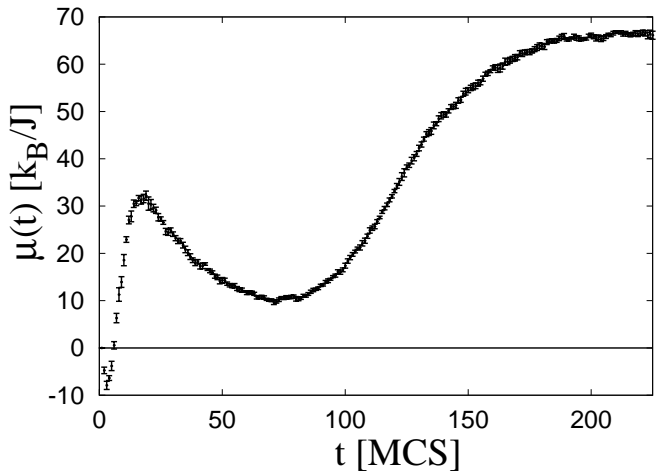}
\caption{Time dependence of the reversed temperature derivative of 
magnetization $\mu(t)$ at $T_{\rm c}=1.442987 J/k_{\rm B}$ for $L=560$.}
\label{fig-mtd-law}
\end{figure}
As explained in the Overview part of this section, the quantity 
$\mu(t) \equiv -{\rm d} \langle |m(t)| \rangle / {\rm d}T$ 
might be suitable for evaluation of $\nu$. 
Using the Hamiltonian of the model (\ref{Hei-Ham}) 
as the energy variable, 
this quantity is expressed as 
\begin{equation}
\mu(L)=-\frac{1}{T^{2}}
\left(\langle |m| {\cal H} \rangle
        - \langle |m| \rangle \langle {\cal H} \rangle \right).
\end{equation}
This expression is nothing but the subtraction between two 
nearly-equal quantities. Although the error bar is expected 
to be large in equilibrium, early-time behaviors from the 
perfectly-disordered state might be different, and law 
data of time dependence of $\mu(t)$ for $L=560$ averaged 
with $80,000$ RNS is displayed in Fig.\ \ref{fig-mtd-law}. 
\begin{figure}
\includegraphics[width=88mm]{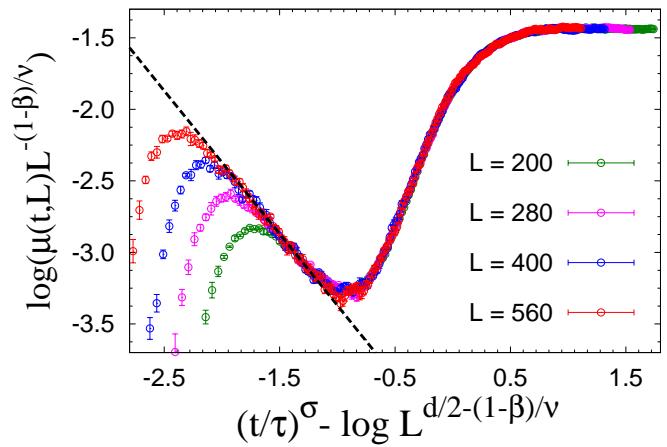}
\caption{(Color online) Nonequilibrium-to-equilibrium 
scaling plot of the quantity $\mu(t,L)$ 
at $T_{\rm c}=1.442987 J/k_{\rm B}$ for $L=200$, 
$280$, $400$ and $560$ with $(1-\beta)/\nu=0.889$, 
$\sigma=0.47$ and $\tau_{\mu}=7.40$. All the data 
except for the initial rapidly-increasing ones seem to 
be located on a single curve, and the dashed line 
stands for the tangent $-1$ as guides for eyes.
}
\label{fig-mtd-sc}
\end{figure}

The quantity $\langle |m| {\cal H} \rangle$ monotonically 
decreases from zero as the simulation time increases, and it 
reaches $\sim - 7 \times 10^{6}$ at $t=225$ MCS for $L=560$. 
That is, $\mu(t)$ is obtained from the subtraction between 
two quantities of order of $\sim 10^{5}$ times larger. 
We may call the error bar of $\mu(t)$ rather ``small" even for 
such severe calculations. This figure also reveals that the time 
dependence of $\mu(t)$ is quite nontrivial: It takes negative 
values in initial several steps, increases rapidly, arrives at a 
maximal value and gradually decreases, shows upturn again 
and finally saturates toward equilibrium. It is not possible to 
decide which region corresponds to the stretched-exponential 
time evolution {\it a prior}, and we have attempted some trials 
and investigated which assumption results in fine scaling 
behaviors in a wide parameter region. 
\begin{figure}
\includegraphics[width=88mm]{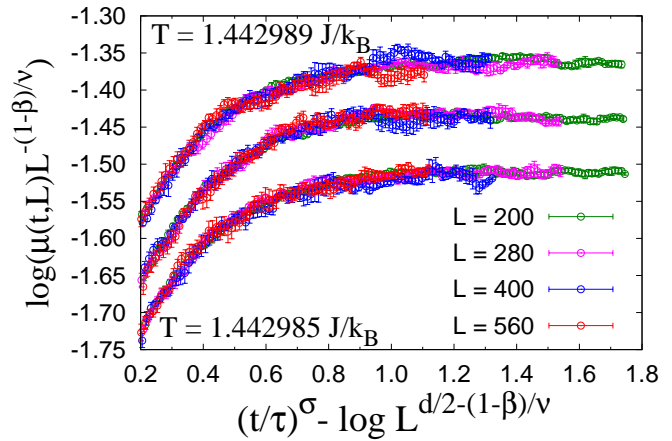}
\caption{(Color online) Enlarged plot in the vicinity of 
equilibrium corresponding to Fig.\ \ref{fig-mtd-sc}  at 
$T=1.442985 J/k_{\rm B}$, $1.442987 J/k_{\rm B}$ and 
$1.442989 J/k_{\rm B}$ with $(1-\beta)/\nu=0.890$, 
$0.889$ and $0.888$, respectively. Error bars are 
a few times larger than symbols, which exhibits 
large fluctuation of the data.
}
\label{fig-mtd-eq}
\end{figure}

We finally find that the stretched-exponential decay 
before the upturn seems reasonable, 
\begin{equation}
\label{mu-sc}
\mu(t,L) \sim L^{d/2} \exp [-(t/\tau_{\mu})^{\sigma}], 
\end{equation}
where the size dependence is derived by multiplying 
random-walk growth of clusters and bulk response. 
From Eq.\ (\ref{mu-sc}) and the size dependence 
in equilibrium, $\mu_{\rm c}(L) \sim L^{(1-\beta)/\nu}$, 
we have the following scaling form, 
\begin{equation}
\log ( \mu(t,L) L^{-(1-\beta)/\nu})
\sim 
(t / \tau_{\mu})^{\sigma} -\log L^{d/2-(1-\beta)/\nu}.
\end{equation}
The scaling plot at $T_{\rm c}=1.442987 J/k_{\rm B}$ with 
$(1-\beta)/\nu=0.889$, $\sigma=0.47$ and $\tau_{\mu}=7.40$ 
is displayed in Fig.\ \ref{fig-mtd-sc}, where all the data 
points except for the initial rapidly-increasing ones for 
$L=200$, $280$, $400$ and $560$ seems to be located 
on a single curve. Actually, the stretched-exponential 
scaling region is rather narrow and precise evaluation 
of $\tau_{\mu}$ is difficult from these data. 
Then, we fit this parameter with the data after the upturn, and 
confirm the acceptable stretched-exponential scaling with this 
estimate afterwards. The data in the vicinity of equilibrium is 
shown in Fig.\ \ref{fig-mtd-eq}, together with the data for 
$T=1.442985 J/k_{\rm B}$ (with $(1-\beta)/\nu=0.890$) 
and $1.442989 J/k_{\rm B}$ (with $(1-\beta)/\nu=0.888$). 
Discrepancy from such scaling behaviors becomes 
clear by changing $(1-\beta)/\nu$ with $\pm 0.005$, 
and we have $(1-\beta)/\nu=0.889(6)$. Combining 
this estimate with Eq.\ (\ref{Tc-est}), we arrive at 
\begin{equation}
\label{nu-est}
1/\nu=1.404 \pm 0.008\ \ {\rm or}\ \ 
\nu=0.712 \pm 0.004.
\end{equation}
Furthermore, when this estimate is coupled with the 
hyperscaling relation, $2-\alpha=d\nu$, we have
\begin{equation}
\label{alpha-est}
\alpha = -0.136 \pm 0.012 \ \ {\rm or}\ \ 
\alpha/\nu = -0.192 \pm 0.016.
\end{equation}

\subsection{Complementary evaluation of $\alpha/\nu$ 
from specific heat and $\nu$ from scaling relation}
\begin{figure}
\includegraphics[width=88mm]{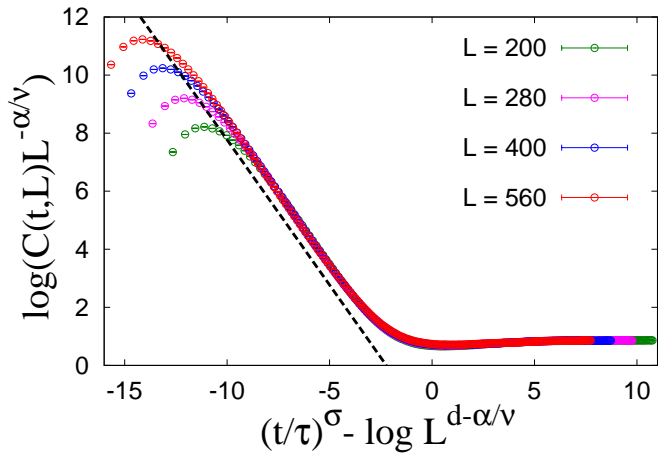}
\caption{(Color online) Nonequilibrium-to-equilibrium 
scaling plot of specific heat at $T_{\rm c}=1.442987 J/k_{\rm B}$ 
for $L=200$, $280$, $400$ and $560$ with $\alpha/\nu=0.074$, 
$\sigma=0.47$ and $\tau_{C}=0.215$. 
All the data except for initial several steps and those during 
the upturn seem to be located on a single curve, and the 
dashed line stands for the tangent $-1$ as guides for eyes.
}
\label{fig-sph-sc}
\end{figure}

The specific heat is expressed as 
\begin{equation}
C(L)=\frac{1}{L^{d}T^{2}}
\left( \langle {\cal H}^{2} \rangle - \langle {\cal H} \rangle^{2} \right).
\end{equation}
This expression is also the subtraction between two 
nearly-equal quantities, of order of $\sim 10^{7}$ times 
larger than $C(L)$ itself in the vicinity of equilibrium for 
$L=560$. Although this ratio is even larger than that of 
$\mu(L)$, error bars of $C(L)$ are smaller than those of 
$\mu(L)$, because fluctuation of energy is much smaller 
than that of magnetization. Although time dependence of 
this quantity is essentially the same as that of $\mu(t)$ in 
Fig.\ \ref{fig-mtd-law}, the stretched-exponential decay of this 
quantity is quite apparent (as will be shown in Fig.\ \ref{fig-sph-sc}), 
\begin{equation}
\label{C-se}
C(t,L) \sim L^{d} \exp \left[ -(t/\tau_{C})^{\sigma} \right],
\end{equation}
where the size dependence originates from bulk response. 
Combining this scaling form with the equilibrium finite-size 
scaling $C_{\rm c}(L)\sim L^{\alpha/\nu}$, we arrive at 
\begin{equation}
\label{C-neq-sc}
\log ( C(t,L) L^{-\alpha/\nu})
\sim 
(t / \tau_{C})^{\sigma} -\log L^{d-\alpha/\nu}.
\end{equation}

\begin{figure}
\includegraphics[width=88mm]{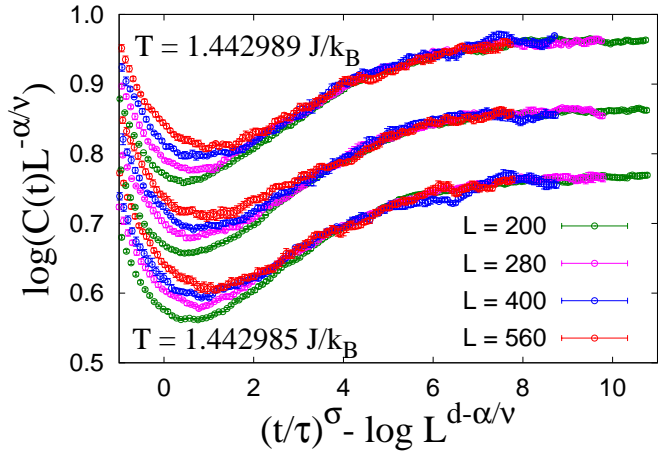}
\caption{(Color online) Enlarged plot in the vicinity 
of equilibrium corresponding to Fig.\ \ref{fig-sph-sc} 
at $T=1.442985 J/k_{\rm B}$, $1.442987 J/k_{\rm B}$ 
and $1.442989 J/k_{\rm B}$ with $\alpha/\nu=0.077$, 
$0..074$ and $0.070$, respectively. 
Error bars are a bit larger than symbols. This figure 
exhibits fluctuation of the data near equilibrium and 
large deviation of the data during the upturn.
}
\label{fig-sph-eq}
\end{figure}

The scaling plot at $T_{\rm c}=1.442987 J/k_{\rm B}$ with 
$\alpha/\nu=0.074(5)$, $\sigma=0.47$ and $\tau_{C}=0.215$ 
for $L=200$, $280$, $400$ and $560$ is given in Fig.\ \ref{fig-sph-sc}. 
The enlarged data after the upturn is shown in Fig.\ \ref{fig-sph-eq} 
together with the data for $T=1.442985 J/k_{\rm B}$ and 
$1.442989 J/k_{\rm B}$ with $\alpha/\nu=0.077(6)$ and 
$0.070(5)$, respectively. 
Error bars of the data are a bit larger than symbols, 
and error bars of estimates of the critical exponent 
$\alpha/\nu$ are also larger than those in other 
exponents. Combining these results we have 
$\alpha/\nu=0.074 \pm 0.009$, and 
$\nu=0.651 \pm 0.002$ from the hyperscaling relation. 
This estimate is not at all consistent with the one 
in Eq.\ (\ref{nu-est}), and such discrepancy had 
already been discussed in the 3D Heisenberg 
model~\cite{Holm93,Brown96,Holm97,Kim00,Brown06}.

\begin{figure}
\includegraphics[width=88mm]{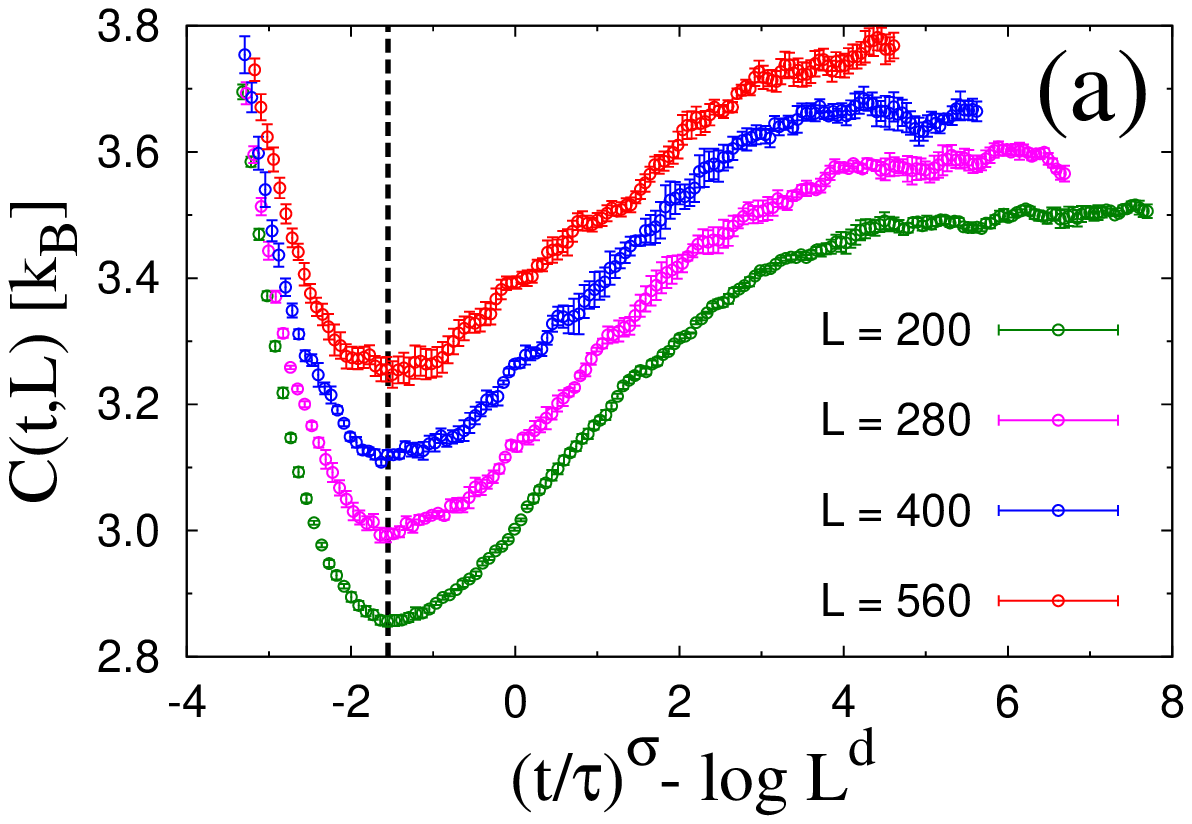}
\includegraphics[width=88mm]{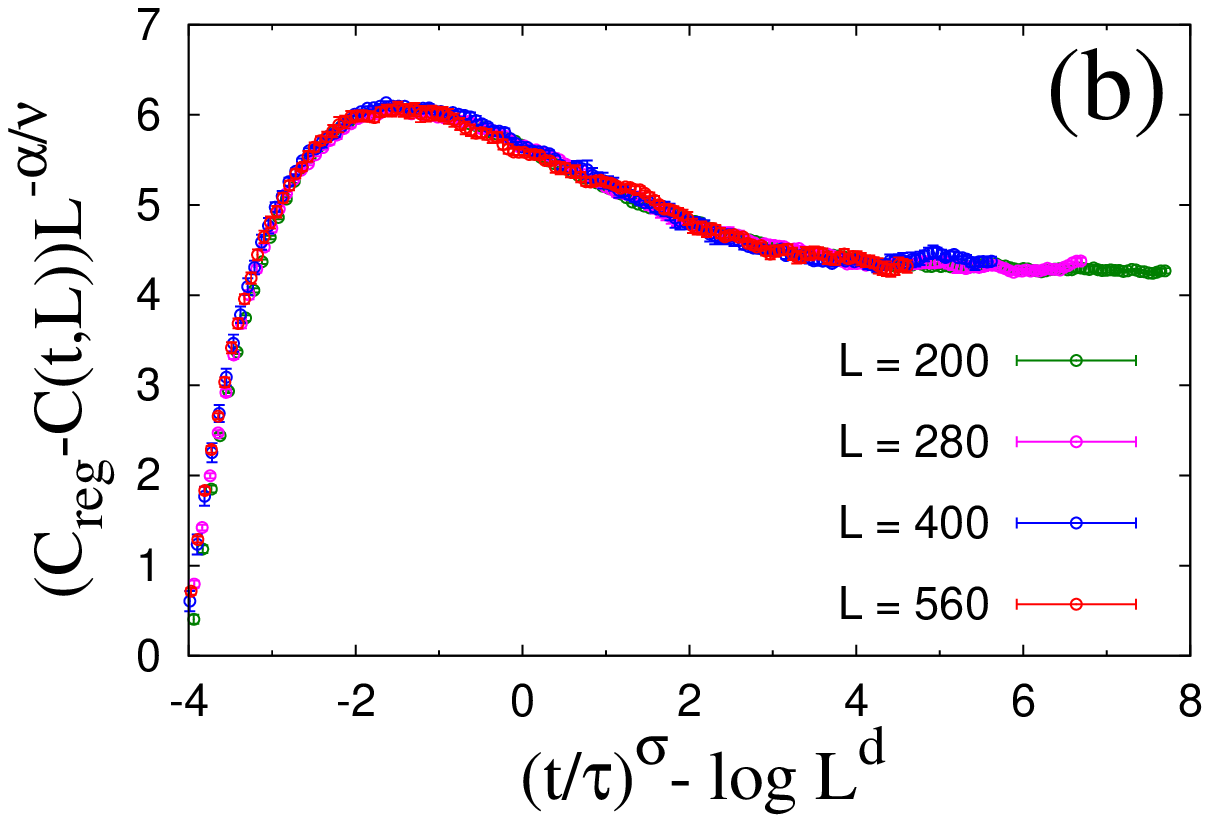}
\caption{(Color online) 
(a) Specific heat at $T_{\rm c}=1.442987 J/k_{\rm B}$ for 
$L=200$, $280$, $400$ and $560$ plotted versus the scaled 
simulation time $t_{\ast} \equiv (t/\tau_{C})^{\sigma}-\log L^{d}$ 
derived from the stretched-exponential scaling form (\ref{C-se}). 
The parameter $\tau_{C} \approx 0.27$ is determined as 
the minimum of the specific heat becomes independent 
of $L$ (emphasized by the dashed line). (b) Scaling of the 
specific heat including the regular part (\ref{C-reg-sc}). 
Assuming $\alpha/\nu=-0.192$ obtained from the scaling 
behavior of $\mu(t)$ in Eq.\ (\ref{alpha-est}) and tuning 
$C_{\rm reg} \approx 5.05 k_{\rm B}$, the coefficient of 
the scaling part of the specific heat becomes independent 
of $L$ in the whole region satisfying $C_{\rm reg} > C(t,L)$.
} 
\label{fig-reg-sc}
\end{figure}

In equilibrium simulations, such a puzzle of 
the specific heat is known to be resolved by 
dividing it to the regular and scaling parts as 
$C(L) = C_{\rm reg} - a L^{\alpha/\nu}$~\cite{Holm93}. 
In the present case, we should also consider time 
dependence of $C(L)$, namely the stretched-exponential 
scaling given in Eq.~(\ref{C-se}). Now we interpret this form 
as an alternative definition of the scaled simulation time 
$t_{\ast} \equiv (t/\tau_{C})^{\sigma}-\log L^{d}$, and the 
parameter $\tau_{C}$ is determined as the minimum of 
$C(t,L)$ to be independent of system size as shown 
in Fig.\ \ref{fig-reg-sc}(a), which corresponds to 
$\tau_{C} \approx 0.27$.

Now we assume that the regular part of the specific heat is 
independent of simulation time $t$, while the coefficient of 
the scaling part depends on $t$ as 
\begin{equation}
\label{C-reg-sc}
C(t,L) = C_{\rm reg} - a(t) L^{\alpha/\nu}.
\end{equation}
The point of this formulation is that $C_{\rm reg}>C(t,L)$ 
or $a(t)>0$ is satisfied in the scaling region, and that the 
exponent $\alpha$ can be negative even though $C(t,L)$ 
increases as the linear size $L$ increases. 
Although it is possible to estimate $C_{\rm reg}$ and $\alpha/\nu$ 
from the scaling plot based on Eq.~(\ref{C-reg-sc}) in principle, 
error bars of the estimates must be much larger than that of 
Eq.~(\ref{alpha-est}), because there exists an extra fitting 
parameter $C_{\rm reg}$ while the fluctuation of the data 
is comparable to that of $\mu(t,L)$. 

Then, we only evaluate $C_{\rm reg}$ by assuming the 
exponent given in Eq.~(\ref{alpha-est}) and confirm the 
validity of the scaling (\ref{C-reg-sc}). In Fig.\ \ref{fig-reg-sc}(b), 
the quantity $(C_{\rm reg}-C(t,L))L^{-\alpha/\nu}$ 
is plotted versus the scaled simulation time 
with $C_{\rm reg} \approx 5.05 k_{\rm B}$ 
at $T_{\rm c}=1.442987 J/k_{\rm B}$ for $L=200$, $280$, 
$400$ and $560$, namely the $t_{\ast}$-dependence 
of the coefficient of the scaling part $a(t)$. 
Apparently, the scaling behavior in Fig.\ \ref{fig-reg-sc}(b) 
seems much better than that in Fig.\ \ref{fig-sph-eq}, 
or the improved scaling form (\ref{C-reg-sc}) with 
negative $\alpha$ might be better than the naive 
scaling form (\ref{C-neq-sc}) with positive $\alpha$.

\section{Discussion}
First, combining the scaling relation 
$\alpha+2\beta+\gamma=2$ and the 
hyperscaling relation, we have 
$2 \beta / \nu + \gamma / \nu=d$. 
On the other hand, from Eqs.\ (\ref{Tc-est}) 
and (\ref{gamma-est}) we result in 
$2\beta/\nu+\gamma/\nu=3.002 \pm 0.012$. 
That is, the critical exponent $\alpha/\nu$ can be 
evaluated from the temperature derivative of the 
magnetization without the hyperscaling relation, 
which is not so trivial as other scaling relations. 

\begin{table}
\begin{tabular}{cclll}
\hline \hline
Refs. & Method & \ \ \ \ \ $\beta$ & \ \ \ \ \ $\gamma$ & \ \ \ \ \ $\nu$ \\
\hline
present I & MC & $0.367(4)$ & $1.404(9)$ & $0.712(4)$ \\
present II & MC & $0.367(2)$ & $1.403(6)$ & $0.712(3)$ \\
\cite{Holm93} (1993) & MC & $0.362(3)$ & $1.389(14)$ & $0.704(6)$ \\
\cite{Hasenbusch01} (2001) & MC & $0.3685(11)$ & $1.393(4)$ & $0.710(2)$ \\
\cite{Campostrini02} (2002) & MC & $0.3691(6)$ & $1.3957(22)$ & $0.7113(11)$ \\
\cite{Campostrini02} (2002) & MC+HTE & $0.3689(3)$ & $1.3960(9)$ & $0.7112(5)$ \\
\cite{Butera97} (1997) & HTE (sc) \ \ & $0.3710(13)$ & $1.406(3)$ & $0.716(2)$ \\
\cite{Butera97} (1997) & HTE (bcc) & $0.3700(13)$ & $1.402(3)$ & $0.714(2)$ \\
\cite{Yukalov98} (1998) & $\epsilon$-Exp. 
& $0.367$ & $1.39$ & $0.708$ \\
\cite{Guida98} (1998) & $\epsilon$-Exp. 
& $0.3655(35)$ & $1.382(9)$ & $0.7045(55)$ \\
\cite{Guida98} (1998) & $d=3$ Exp. 
& $0.3662(25)$ & $1.3895(50)$ & $0.7073(35)$ \\
\cite{Jasch01} (2001) & $d=3$ Exp. 
& $0.3655(5)$ & $1.3882(10)$ & $0.7062(7)$ \\
\hline \hline
\end{tabular}
\caption{
List of critical exponents estimated in the present study 
and in several previous numerical studies.
}
\label{tab1}
\end{table}
Next, we compare the estimates of $\beta$, $\gamma$ and $\nu$
obtained from Eqs.\ (\ref{Tc-est}), (\ref{gamma-est}) and (\ref{nu-est})
(present I) with several previous studies based on the 
Monte Carlo method~\cite{Holm93,Hasenbusch01,Campostrini02}, 
high-temperature expansion~\cite{Campostrini02,Butera97}, 
$\epsilon$-expansion~\cite{Yukalov98,Guida98} or 
$d=3$ expansion~\cite{Guida98,Jasch01} in Table \ref{tab1}. 
More comprehensive list of references 
was given in Ref.~\cite{Campostrini02}. 
Note that our estimates shown above are based on ``safe" evaluation.  
That is, the critical exponents are determined in order to cover the 
whole possible range of the critical scaling. If we assume that 
$T_{\rm c}$ locates in the middle of the scaling region and the critical 
exponents are evaluated just at $T_{\rm c}=1.442987(1) J/k_{\rm B}$, 
reasonable reduction of error bars is possible (present II).

From this table we may at least conclude that the estimates of 
critical exponents in the present study are consistent with those 
in previous studies, even though error bars are not so small. 
The dominant origin of statistical errors in our estimates is 
large fluctuations of $\mu(t,L)$. Since all the critical exponents 
evaluated with finite-size scaling analyses are scaled by $\nu$, 
Precision of this exponent is crucial. Nevertheless, the main 
purpose of the present study is to confirm the nontrivial 
stretched-exponential critical scaling in the 3D Heisenberg 
model, and evaluation of the critical exponents is not more 
than consistency check of this alternative scheme.

Actually, information of the critical exponents is not included in 
the stretched-exponential critical scaling of physical quantities 
in itself. Such information is introduced from the finite-size 
scaling of physical quantities in equilibrium, and the 
nonequilibrium-to-equilibrium scaling scheme 
enables to extract such information from the 
off-scaling data prior to equilibrium behaviors. 
The most promising application of the present scheme 
is to characterize phase transitions with nonequilibrium 
behaviors. In the standard NER scheme based on the 
local-update algorithms, power-law behaviors take place 
in general even in the BKT~\cite{Ozeki-BKT} or 
first-order~\cite{Ozeki-1st} phase transitions, and such 
characterization was made in a different manner by 
assuming the nature of phase transitions in advance. 
The situation might be different in the present cluster 
NER scheme, and investigations along this direction 
is now in progress~\cite{Nono-letter}.

The stretched-exponential nonequilibrium critical scaling initially 
found in the 2D Ising model~\cite{Nonomura14} is now conformed 
in the 3D Heisenberg model. The exponent of stretched-exponential 
time dependence $\sigma \approx 1/2$ seems consistent with that of 
the 3D XY model~\cite{Nono-letter}, while not consistent with that of the 
2D or 3D Ising models, $\sigma \approx 1/3$~\cite{Nonomura14,Nono-full}. 
Even though our simulations on 3D vector spin models are based 
on the ``embedded-Ising-spin" formalism, the exponent $\sigma$ 
is different from that of the Ising models. That is, this exponent is 
not determined by mere formalism, but by fundamental physical 
properties. Further details of such universal behaviors of the 
exponent $\sigma$ will be explained elsewhere~\cite{Nono-full}.

\section{Summary}
Critical behaviors of the 3D Heisenberg model are investigated 
with the nonequilibrium ordering from the perfectly-disordered 
state in the Swendsen-Wang (SW) algorithm characterized 
by the stretched-exponential critical scaling form. 
Calculations from the perfectly-ordered state in the SW algorithm 
result in larger initial-time deviation and slower off-critical deviation, 
and those from the perfectly-disordered state in the Wolff algorithm 
show a power-law behavior similar to that in local-update algorithms. 
From simulations based on the ``embedded-Ising spin" formalism 
we have the nonequilibrium-to-equilibrium scaling similarly to the 
3D XY model, and all the data for $L=200$, $280$, $400$ and 
$560$ seem to be on a single curve in the whole simulation-time 
region with the stretched-exponential exponent $\sigma=0.47(1)$. 

Precise values of the critical temperature and the critical exponent 
$\beta/\nu$ are evaluated simultaneously from the scaling analysis 
of the magnetization. Similarly, the critical exponent $\gamma/\nu$ 
is obtained from the scaling analysis of the magnetic susceptibility. 
The critical exponent $\nu$ is evaluated from the scaling analysis 
of the temperature derivative of magnetization, and the 
critical exponent $\alpha <0$ can be derived from the 
hyperscaling relation. A consistent estimate of $\alpha/\nu$ can be 
obtained from the specific heat by assuming the scaling form with 
the size-independent regular part and the negative scaling part. 

Although information on standard critical behaviors such as the 
critical exponents is not included in the stretched-exponential 
critical scaling behaviors, such information can be extracted 
from off-scaling behaviors with the nonequilibrium-to-equilibrium 
scaling scheme, and the critical exponents comparable to 
previous studies can be obtained from the present framework 
without full equilibration. 

\begin{acknowledgments}
The random-number generator 
MT19937~\cite{MT} was used for numerical calculations.
Part of the calculations were performed 
on Numerical Materials Simulator at NIMS. 
\end{acknowledgments}

\bigskip

\appendix

\section{Evaluation of critical temperature from single system}
\subsection{Double-log plot}
The simplest way to evaluate the critical temperature $T_{\rm c}$ 
from the early-time dependence of magnetization (\ref{se-ord}) 
or (\ref{se-dis}) is the double-log plot. When both sides 
of these equations are divided by the coefficient of rhs 
(expressed as $C$) and take logarithm twice, we have 
\begin{equation}
\log ( \log ( \langle |m(t)| \rangle / C ) ) \sim \mp \sigma \log ( t /\tau_{m}).
\end{equation}
Since Eq.\ (\ref{se-ord}) is obtained by relaxation from the 
perfectly-ordered state, the coefficient $C$ is of order of unity, 
and a naive plot assuming $C \equiv 1$ may be satisfactory 
enough. On the other hand, Eq.\ (\ref{se-dis}) is obtained by 
ordering from the perfectly-disordered state, and the small 
coefficient $C$ should be taken into account explicitly. 

\begin{figure}
\includegraphics[width=88mm]{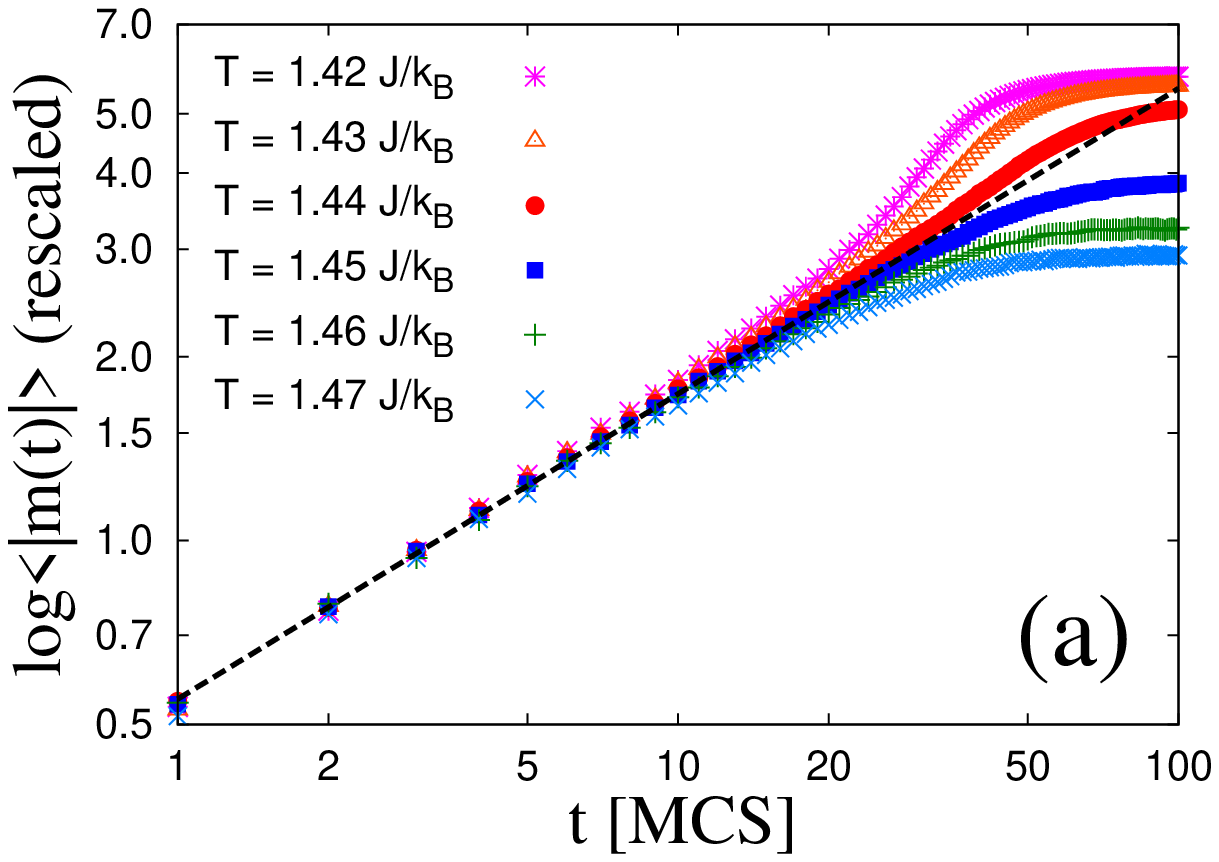}
\includegraphics[width=88mm]{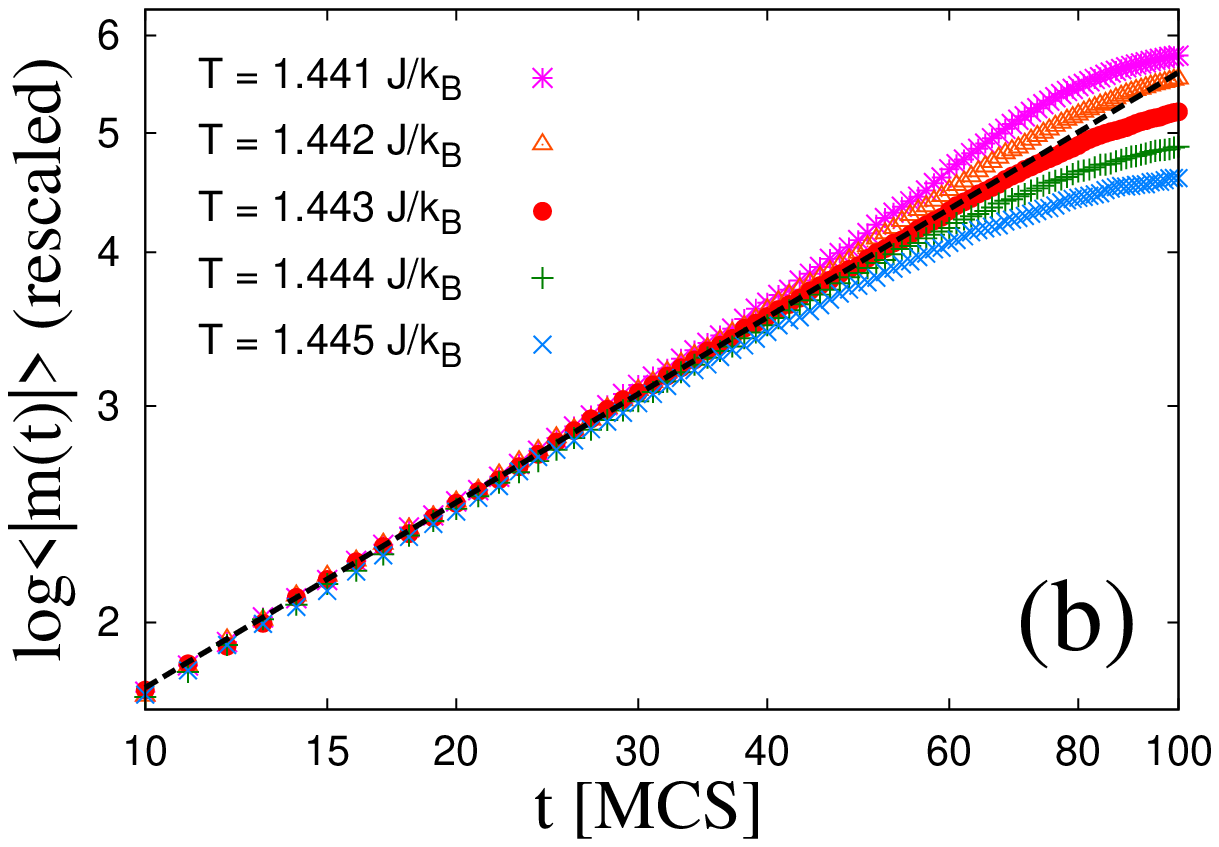}
\caption{(Color online) Double-log plot of magnetization 
in the SW algorithm from the perfectly-disordered state 
for various temperatures around $T_{\rm c}$ for 
(a) $L=100$ and (b) $L=200$. The dashed line 
corresponds to the stretched-exponential 
ordering with $\sigma=0.5$.
}
\label{fig-dl}
\end{figure}

Practically, the unknown parameter $C$ can be determined 
by aligning the early-time data on a straight line, as shown in 
Fig.\ \ref{fig-dl}(a) with the data for $L=100$ with $2,000$ RNS 
in the SW algorithm from the perfectly-disordered state. 
Apparently, the ordering process in the double-log plot is 
faster or slower than linear for $T \leq 1.44 J/k_{\rm B}$ 
or $T \geq 1.45 J/k_{\rm B}$, respectively, which means 
$1.44 J/k_{\rm B} < T_{\rm c} < 1.45 J/k_{\rm B}$. 
Tangent of the dashed line is $0.5$, which 
is consistent with $\sigma \approx 0.5$. 
Then, the variance of temperatures is reduced of one order in 
Fig.\ \ref{fig-dl}(b) with the data for $L=200$ with $2,000$ RNS, 
which results in $T_{\rm c} \approx 1.443 J/k_{\rm B}$ 
with $\sigma \approx 0.5$.

Differently from the standard NER scheme based on the 
power-law behavior of physical quantities, the present 
scheme requires the rescaling of physical quantities 
with the unknown coefficient of stretched-exponential 
behaviors, and therefore precise evaluation 
of $T_{\rm c}$ is difficult. 

\subsection{Semi-log plot with fixed $\sigma$}
\begin{figure}
\includegraphics[width=88mm]{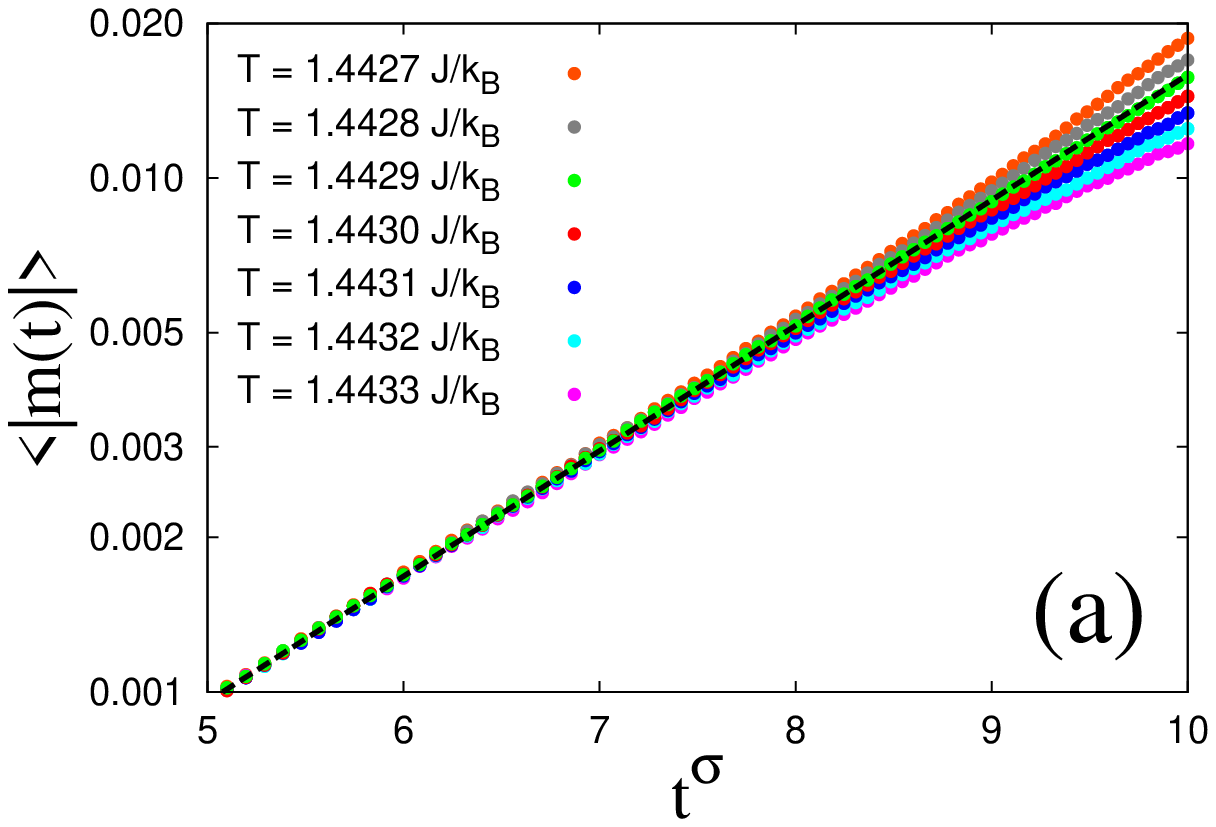}
\includegraphics[width=88mm]{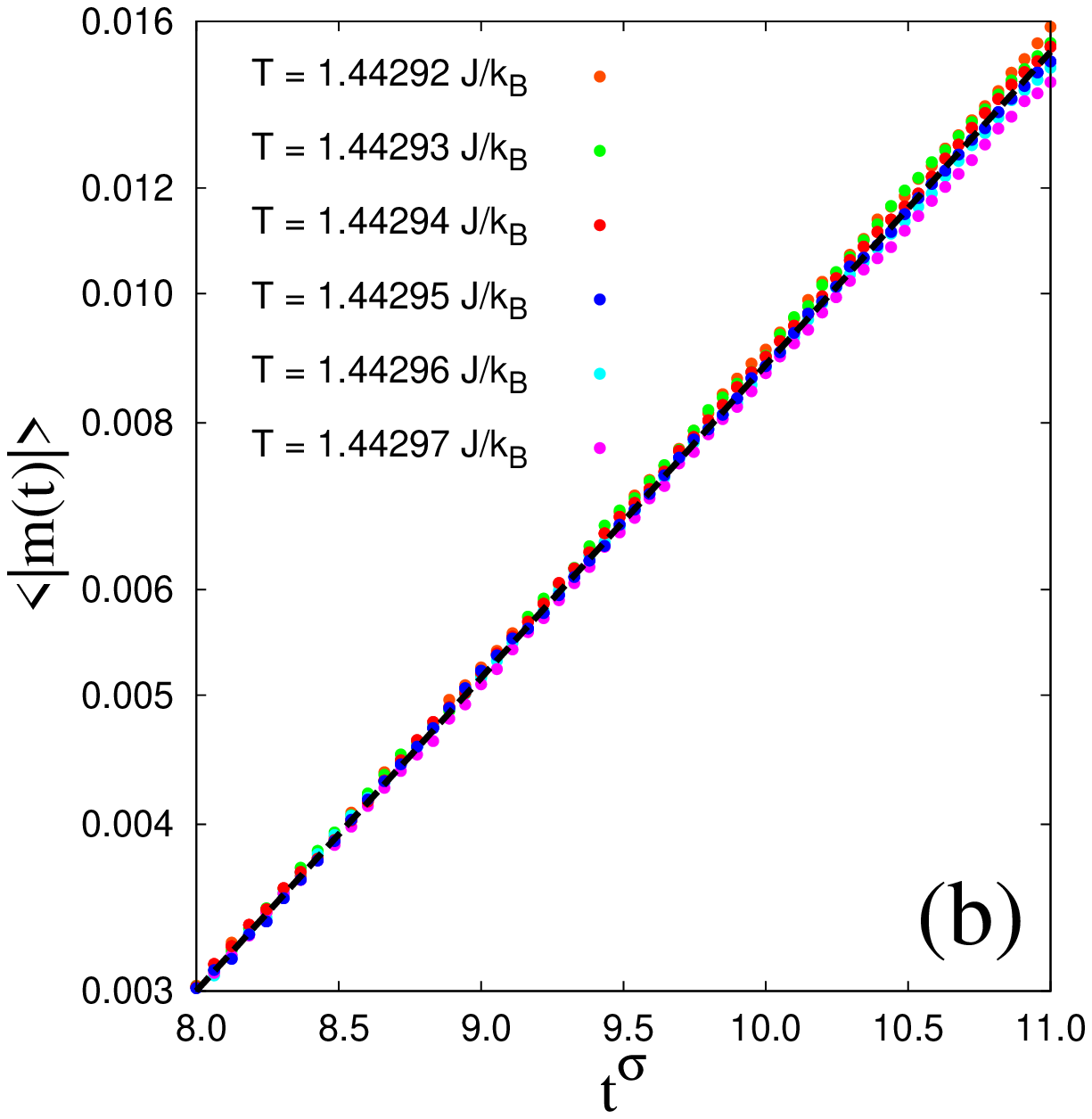}
\caption{(Color online) Semi-log plot of magnetization 
versus $t^{\sigma}$ with $\sigma \equiv 0.5$ in the 
SW algorithm from the perfectly-disordered state 
for various temperatures around $T_{\rm c}$ for 
(a) $L=560$ and (b) $L=800$. The dashed line 
corresponds to $T_{\rm c}$ obtained from these 
figures. Discrepancy with the estimate in Section III 
reveals the limitation of this graphical scheme.
}
\label{fig-sl}
\end{figure}

If we intend to estimate more accurate $T_{\rm c}$ from 
the double-log plot, we may determine the linearity of the 
early-time data assuming the exponent $\sigma \equiv 0.5$. 
However, as long as $\sigma$ is fixed {\it a prior}, plotting 
$\log \langle |m(t)| \rangle$ versus $t^{\sigma}$ is more 
straightforward and precise, because no adjusting 
parameter other than $\sigma$ is included. 
Such semi-log plot with $\sigma \equiv 0.5$ is shown 
in Fig.\ \ref{fig-sl}(a) for $L=560$ with $10,000$ RNS 
around $T=1.443 J/k_{\rm B}$. Evaluation of $T_{\rm c}$ 
is not an easy task because the data start deviating 
at $t = 50 \sim 60$ MCS, while they already begin to 
saturate at $t = 70 \sim 80$ MCS. Observing the data 
on the onset of deviation carefully, we may conclude 
$1.4429 J/k_{\rm B}<T_{\rm c}< 1.4430 J/k_{\rm B}$. 

Then, the variance of temperatures is reduced of one more 
order in Fig.\ \ref{fig-sl}(b) with the data for $L=800$ with 
$5,000$ RNS. Although the data seem to suggest 
$1.44294 J/k_{\rm B}<T_{\rm c}<1.44295 J/k_{\rm B}$, 
this estimate of $T_{\rm c}$ is not consistent with the 
scaling plot given in Eq.\ (\ref{eq-scale}) at all. 
Even though the data starts deviating at 
$t = 80 \sim 100$ MCS, they already begin 
to saturate in such simulation-time region. 
Downward-bending behavior due to saturation 
and upward-bending behavior below $T_{\rm c}$ 
are cancelled with each other in this region, and 
therefore $T_{\rm c}$ is underestimated when it is 
determined by the linearity of the nonequilibrium data.
The scaling analysis explained in Section III 
is more reasonable and precise.

\section{Nonequilibrium relaxation from the perfectly-ordered state}
\subsection{Numerical results in the SW algorithm}
\begin{figure}
\includegraphics[width=88mm]{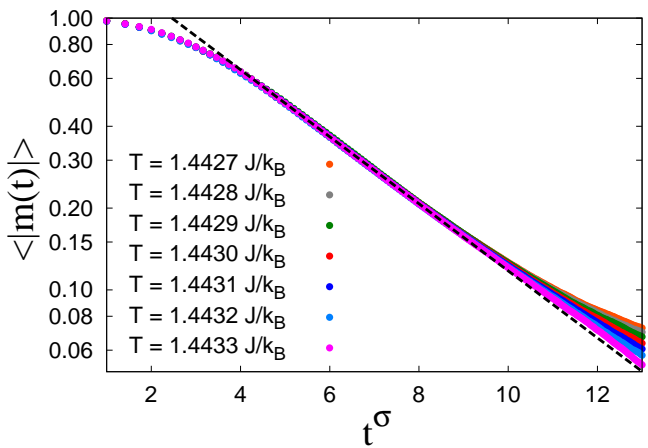}
\caption{(Color online) Semi-log plot of magnetization 
versus $t^{\sigma}$ with $\sigma \equiv 0.5$ in the SW 
algorithm from the perfectly-ordered state for various 
temperatures around $T_{\rm c}$ for $L=560$. 
The dashed line emphasizes the region in which 
the stretched-exponential decay holds well, and 
it clarifies early-time rounding and saturation of data 
earlier than deviation induced by varying temperatures.
}
\label{fig-SWF}
\end{figure}
Relaxation data of magnetization from the perfectly-ordered 
state in the SW algorithm for $L=560$ with $400$ RNS are 
displayed in Fig.\ \ref{fig-SWF} with a semi-log plot versus 
$t^{\sigma}$, $\sigma \equiv 0.5$ similarly to Fig.\ \ref{fig-sl}. 
This figure shows merits and demerits of this scheme. First, 
although average number of RNS is one order smaller than 
that in Fig.\ \ref{fig-sl}, diversity of the data looks comparable. 
On the other hand, stable stretched-exponential decay only 
starts from $t \sim 20$ MCS as emphasized by the dashed 
line, while such behavior is observed from the beginning in 
Fig.\ \ref{fig-dl}. Moreover, the onset of deviation of the data 
induced by variance of temperature is $t \sim 80$ MCS, 
which is later than that in Fig.\ \ref{fig-sl}, and even 
later than the onset of saturation in Fig.\ \ref{fig-SWF}, 
$t \sim 70$ MCS, which is comparable to that in 
Fig.\ \ref{fig-sl}. That is, evaluation of $T_{\rm c}$ from 
deviation of the data does not make sense, because 
this deviation is already affected by saturation behavior.

\subsection{Numerical results in the Wolff algorithm}
\begin{figure}
\includegraphics[width=86mm]{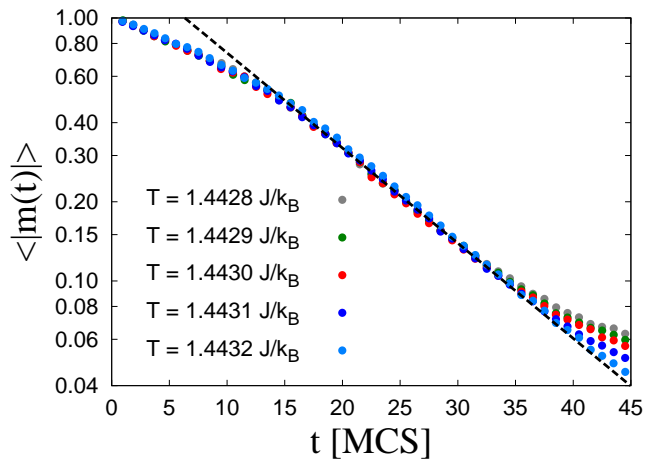}
\caption{(Color online) Semi-log plot of magnetization 
in the Wolff algorithm from the perfectly-ordered state 
for various temperatures around $T_{\rm c}$ for $L=560$. 
The dashed line emphasizes the region in which the 
simple exponential decay holds well. Although time 
dependence looks similar to that in Fig.\ \ref{fig-SWF}, 
we take $\sigma \equiv 1$ in this figure.
}
\label{fig-WolF}
\end{figure}
Relaxation data of magnetization from the perfectly-ordered state 
in the Wolff algorithm for $L=560$ with $100$ RNS are shown 
in Fig.\ \ref{fig-WolF} with a standard semi-log plot versus $t$, 
or the critical magnetization decays exponentially in this case. 
Similarly to the corresponding relaxation in the SW algorithm, 
stable exponential decay only starts from $t \sim 20$ MCS, and 
saturation of the data begins earlier than deviation induced 
by variance of temperature. Rather small number of RNS 
is due to larger numerical cost than that in the SW algorithm. 
Although required simulation time for equilibration looks much 
smaller than that in the SW algorithm, this time is scaled by 
the flipped cluster size in each step in the Wolff algorithm. 
Since the exponent $\beta/\nu$ is rather large ($\approx 0.5$) 
in the present model, the critical magnetization $m_{\rm c}(L)$ 
rapidly decreases as the linear size $L$ increases. This situation 
is quite different from that in the 2D Ising model~\cite{Nonomura14}, 
where the small exponent $\beta/\nu=1/8$ ensured efficient update 
in the Wolff algorithm.

Aside from numerical efficiency, it is theoretically interesting 
that the exponent of the stretched-exponential decay (simple 
exponential decay can be regarded as the $\sigma=1$ case) 
depends on the species of the cluster algorithm. 
Such behaviors did not take place in the 2D Ising 
model~\cite{Nonomura14}, where the decay 
exponent from the perfectly-ordered state is the same 
in the SW or Wolff algorithms. (Note that ordering behaviors 
from the perfectly-disordered state is quite different in the 
two algorithms even in the 2D Ising model~\cite{Du}.) 
Although the direction of flip is trivial 
in the Ising model, it is chosen randomly in the Heisenberg model. 
In each step the alignment is common in the whole system in the 
SW algorithm, but it is random from cluster to cluster in the Wolff 
algorithm. Absence of correlation between clusters is characteristic to 
behaviors above $T_{\rm c}$, and it results in the exponential decay. 
That is, the critical slowing down does not exist in the relaxation 
from the perfectly-ordered state in the Heisenberg model 
in the Wolff algorithm. Although this behavior is favorable 
for equilibration, it is not suitable for NER analyses.


\begin{thebibliography}{99}
\bibitem{app-order}
Since the scheme explained in Appendix A is used in Appendix B, 
the order of the appendices is reversed with the appearance 
in the main text.
\end{thebibliography}
\end{document}